\documentclass[english,10.5pt]{extarticle}
\usepackage{mathptmx}
\usepackage[T1]{fontenc}
\usepackage[utf8]{inputenc}
\usepackage[a4paper]{geometry}
\usepackage{fancyhdr}
\usepackage{color}
\definecolor{document_fontcolor}{rgb}{0, 0, 0}
\color{document_fontcolor}
\usepackage{babel}
\usepackage{amsmath}
\usepackage{amssymb}
\usepackage{graphicx}
\usepackage[unicode=true,
 bookmarks=false,
 breaklinks=false,pdfborder={0 0 1},backref=section,colorlinks=true]
 {hyperref}
\hypersetup{
 linkcolor=blue}

\makeatletter
\@ifundefined{date}{}{\date{}}
\usepackage[affil-it]{authblk}
\usepackage{etoolbox}
\usepackage{setspace}
\usepackage{amsthm}\usepackage{tikz}
\usepackage{listings}
\usepackage{caption}
\usepackage{placeins}
\usepackage{booktabs}
\usepackage{enumerate}
\usepackage{titlesec}
\usepackage{anyfontsize}
\usepackage{t1enc}
\usepackage{fancyhdr}
\usepackage{float}
\usepackage{xspace}\usepackage{scalefnt}

\usepackage{soul}% to be able to use \hl{} command
\usepackage{tabulary}
\usepackage{textcase}

\usepackage[colorinlistoftodos]{todonotes}

\titlespacing*{\section}
{0pt}{12pt}{0pt}
\titlespacing*{\subsection}
{0pt}{12pt}{0pt}
\titlespacing*{\subsubsection}
{0pt}{12pt}{0pt}
\titleformat{\section}
 {\huge\sffamily\Large\bfseries}
 {\thesection}{1em}{}
   \titleformat{\subsection}[block]{\bfseries}{\thesubsection}{.5em}{}
   \titleformat{\subsubsection}[block]{\bfseries}{\thesubsubsection}{.5em}{}

\titleformat{\section}{\fontsize{12}{19}\bfseries}{\thesection}{1em}{}

\patchcmd{\@maketitle}{\LARGE \@title}{\fontsize{14}{19.2}\selectfont\@title}{}{} % original: 18, 19.2
\makeatother

\begin{document}
\global\long\def\tht{\vartheta}%
\global\long\def\ph{\varphi}%
\global\long\def\balpha{\boldsymbol{\alpha}}%
\global\long\def\btheta{\boldsymbol{\theta}}%
\global\long\def\bJ{\boldsymbol{J}}%
\global\long\def\bGamma{\boldsymbol{\Gamma}}%
\global\long\def\bOmega{\boldsymbol{\Omega}}%
\global\long\def\d{\mathrm{d}}%
\global\long\def\t#1{\text{#1}}%
\global\long\def\m{\text{m}}%
\global\long\def\v#1{\mathbf{#1}}%
\global\long\def\u#1{\underline{#1}}%

\global\long\def\t#1{\mathbf{#1}}%
\global\long\def\bA{\boldsymbol{A}}%
\global\long\def\bB{\boldsymbol{B}}%
\global\long\def\c{\mathrm{c}}%
\global\long\def\difp#1#2{\frac{\partial#1}{\partial#2}}%
\global\long\def\xset{{\bf x}}%
\global\long\def\zset{{\bf z}}%
\global\long\def\qset{{\bf q}}%
\global\long\def\pset{{\bf p}}%
\global\long\def\wset{{\bf w}}%
\global\long\def\ei{{\bf \mathrm{ei}}}%
\global\long\def\ie{{\bf \mathrm{ie}}}%
\global\long\def\pt{\partial}%
 
\global\long\def\no{\nonumber}%
 
\global\long\def\del{\delta}%
 
\global\long\def\tg{\tau_{g}}%
 
\global\long\def\tba{\bar{\tau}}%
\global\long\def\s{\mathrm{s}}%
\global\long\def\a{\mathrm{a}}%
\global\long\def\f{\mathrm{f}}%

\title{\vspace{-2cm}
%\vspace{-6cm}\includegraphics[width=17cm]{ICA_capa2019.png}\\[0.5cm] 
\textbf{Physics-informed transfer path analysis with parameter estimation using Gaussian processes}
	\author[ ]{Christopher ALBERT$^{(1)}$}
    \date{\today}
 	\affil[(1)]{Max-Planck-Institut für Plasmaphysik, 85748 Garching, Germany, albert@alumni.tugraz.at}
}

% \begin{titlepage}
\clearpage\setcounter{page}{1} \maketitle \thispagestyle{empty}
\fancypagestyle{empty}{\fancyhf{} \fancyfoot[L]{Presented
at the 23rd International Congress on Acoustics, Sep 09-13, 2019 in
Aachen, Germany\vspace{-2cm}
}}

\subsection*{\fontsize{10.5}{19.2}Abstract}

{\fontsize{10.5}{60}\selectfont Gaussian processes regression is
applied to augment experimental data of transfer-path analysis (TPA)
by known information about the underlying physical properties of the
system under investigation. The approach can be used as an alternative
to model updating and is also applicable if no detailed simulation
model of the system exists. For vibro-acoustic systems at least three
features are known. Firstly, observable quantities fulfill a wave
equation or a Helmholtz-like equation in the frequency domain. Secondly,
the relation between pressure/stress and displacement/velocity/acceleration
are known via constitutive relations involving mass density and elastic
constants of the material. The latter also determine the propagation
speed of waves. Thirdly, the geometry of the system is often known
up to a certain accuracy. Here it is demonstrated that taking into
account this information can potentially enhance TPA results and quantify
their uncertainties at the same time. In particular this is the case
for noisy measurement data and if material parameters and source distributions
are (partly) unknown. Due to the probabilistic nature of the procedure
unknown parameters can be estimated, making the method also applicable
to material characterization as an inverse problem.}

{\fontsize{11}{60}\selectfont Keywords: transfer path analysis,
field reconstruction, inverse problem} % at least 1 keyword is required (maximum of 5 keywords)

\fontdimen2\font=4pt

\section{\uppercase{Introduction}}

% \section{Introduction}

Data from acoustical measurements in experimental setups are traditionally
used as-is, without taking into account physical laws that restrict
possible outcomes to plausible results. Here it is shown how acoustical
measurements, in particular for TPA, can be made more reliable by
leveraging additional knowledge of the system under investigation.
A common way to put physical information into use consists in the
construction of a simulation model and using model updating for certain
parameters based on experimental measurements. In a purely experimental
setup for TPA this procedure if often too time-consuming to be realized.
As an alternative a meshless probabilistic approach is introduced
here. The acoustic pressure field is modeled as a Gaussian random
field (used synonymously with Gaussian processes here, for more details
see the book of Rasmussen and Williams~\cite{Rasmussen2006}). Gaussian
process (GP) regression is used to update information based on measurements,
thereby obtaining both, expected values and uncertainties for quantities
of interest. Here we will augment GP regression by partial differential
equations representing physical information based on the method pointed
out by van den Boogaart~\cite{van2001kriging}. More recently an
alternative variant has been introduced by Raissi et al.~\cite{Raissi2017}.
In contrast to the latter where sources are modeled as a GP, \cite{van2001kriging}
treats sources as fixed, restricting kernel functions to physically
possible cases for the homogeneous part of the equation. This paper
will give a first taste on the capabilities of this powerful technique
for acoustics.

For many application acoustic fields can be assumed to follow a linear
wave equation. The speed of sound $c$ is either known exactly or
approximately at well-defined environmental conditions. It will be
demonstrated that this is sufficient to reconstruct sound pressure
fields in the vicinity of measurement positions and domains enclosed
by sensors. Numerical experiments and real-world measurements show
that the Nyquist criterion of two microphones per wavelength is sufficient
for an accurate reconstruction in-between. Finally, the estimation
of acoustic source strengths for TPA is explored in a numerical model
of a simplified structure representing an engine bay. Despite the
use of FEM simulations for this it should be emphasized that the reconstruction
relies only on microphone positions and signals together with the
speed of sound $c$. It is therefore directly applicable to purely
experimental setups as well as a surrogate representation of fields
from numerical models. 

\section{\uppercase{Acoustics with Gaussian random fields}\label{sec:Acoustics-with-Gaussian}}

Consider an inviscid fluid without equilibrium flow with possibly
space-dependent equilibrium density $\rho_{0}(\xset)$ and speed of
sound $c(\xset)$. For a sufficiently small time-harmonic perturbation
at circular frequency $\omega$ a pressure perturbation $p(\xset)$
is related to volumetric sources $w(\xset)$ and momentum sources
(body force density) $\v f(\xset)$ via the Helmholtz equation
\begin{equation}
\Delta p+k^{2}p=i\omega\rho_{0}w-\nabla\cdot\v f\equiv q.\label{eq:p}
\end{equation}
Here $k=k(\xset)=\omega/c(\xset)$ is the wavenumber at frequency
$\omega$. We will now leverage this knowledge about $p$ to augment
possibly noisy experimental data from measurements.

To apply GP regression we model $p$ as a Gaussian random field
\begin{equation}
p\sim\mathcal{N}\left(0,K(\xset,\xset^{\prime})\right).\label{eq:gp}
\end{equation}
Generally speaking this means that $p(\xset)$ is the realization
of a normally distributed random variable with zero mean at each point
$\xset$. Pointwise values $p(\xset)$ are however not independent
from values $p(\xset^{\prime})$ at neighbouring positions $\xset^{\prime}$.
For example, if the distance $|\xset-\xset^{\prime}|$ between two
points is much smaller than a wavelength, local pressure values will
be nearly the same. This fact is taken into account by introducing
correlations via a \emph{covariance function }$K(\xset,\xset^{\prime})$
that is also called a \emph{kernel}. 

In addition to $\xset$ and $\xset^{\prime}$ the kernel usually depends
on hyperparameters, in particular the correlation length, that can
be either fixed or estimated. Despite the assumption of a normal distribution
in $p$ at each point, the kernel $K(\xset,\xset^{\prime})$ does
not have to be a Gaussian in $|\xset-\xset^{\prime}|$ (in that special
case $K$ is called a \emph{squared exponential kernel}). In fact,
application of GP regression to unstructured data usually involves
choosing an appropriate kernel $K$ based on the data themselves.
The case is however different for the application to quantities such
as acoustic pressure $p$ that are known to fulfil laws of physics
formulated as differential equations. In particular the wave equation~\eqref{eq:p}
puts constraints on the possible choice of the kernel function $K$
for a Gaussian process modeling $p$. In source-free regions we have
the homogeneous equation
\begin{equation}
p=-k^{-2}\Delta p.\label{eq:psf}
\end{equation}
Application of the Laplacian together with the scaling by $-k^{2}$
yields also a Gaussian field $\Delta p\sim\mathcal{N}\left(0,L(\xset,\xset^{\prime})\right)$
with a new kernel $L$. The latter is computed from the differential
operator acting in both, $\xset$ and $\xset^{\prime}$ (see \cite{Rasmussen2006},
p.~191 and references in~\cite{van2001kriging}), so
\begin{align}
L(\xset,\xset^{\prime}) & =k^{-2}(\xset)k^{-2}(\xset^{\prime})\Delta_{\xset}\Delta_{\xset^{\prime}}K(\xset,\xset^{\prime}).
\end{align}
To be able to fulfill the homogeneous wave equation~\eqref{eq:psf},
both sides should follow the same distribution. Thus we require kernels
$L=K$ to coincide. This means that $K$ must be an eigenfunction
of the Laplacian in both, $\xset$ and $\xset^{\prime}$, with the
product of eigenvalues being $k^{2}(\xset)k^{2}(\xset^{\prime})$.
Due to lack of directional information for we consider isotropic \emph{radial
basis functions }$K(\xset,\xset^{\prime})=K(|\xset-\xset^{\prime}|)$.
Under the assumption of normalized data we set $K_{p}(0)=1$ meaning
a local variance of $1$. If we restrict ourselves to a real (no damping)
and spatially constant speed of sound $c$ and therefore also wavenumber
$k$, we can immediately compute the relevant eigenfunctions. For
1D problems with $\xset=x$ we obtain
\begin{equation}
K^{1\mathrm{D}}(x,x^{\prime})=\cos(k(x-x^{\prime})).\label{eq:1D}
\end{equation}
For 2D problems with $\xset=(x,y)$ we have the Bessel function
\begin{equation}
K^{2\mathrm{D}}(\xset,\xset^{\prime})=J_{0}(k|\xset-\xset^{\prime}|),
\end{equation}
and for 3D problems with $\xset=(x,y,z)$ the kernel is a spherical
Bessel (sinc) function
\begin{equation}
K^{3\mathrm{D}}(\xset,\xset^{\prime})=j_{0}(k|\xset-\xset^{\prime}|)=\frac{\sin(k|\v x-\v x^{\prime}|)}{|\v x-\v x^{\prime}|}.\label{eq:3D}
\end{equation}
The close relation to the usual Helmholtz kernels, proportional to
complex exponentials in 1D and 3D and a Hankel function in 2D, is
apparent. The wavenumber $k$ is a hyperparameter, describing correlation
at the scale of the wavelength $\lambda=2\pi/k$. For the following
analysis $k$ will be given exactly from $\omega$ and $c$. The 1D
kernel $K^{1\mathrm{D}}$ yields complete periodic correlation of
$p$ at distances of full wavelengths $n\lambda$ with integer $n$
and anticorrelation at half wavelengths $(n+1/2)\lambda$. In a physical
interpretation this is related to the fact that the field is confined
to a line from which no information can escape to the side. In contrast,
for 2D and 3D problems the oscillating kernels decay in amplitude
at increasing distance $|\xset-\xset^{\prime}|$ due to geometric
spreading of the according field. 

\begin{figure}

\begin{minipage}[t]{0.5\columnwidth}%
\begin{center}
\includegraphics[width=1\textwidth]{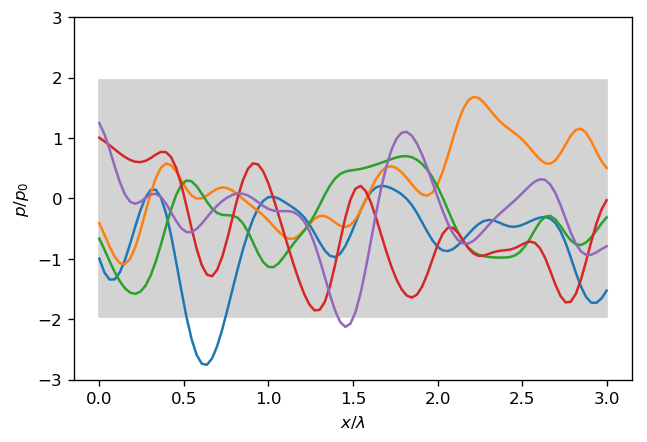}\\
(a)
\par\end{center}%
\end{minipage}%
\begin{minipage}[t]{0.5\columnwidth}%
\begin{center}
\includegraphics[width=1\textwidth]{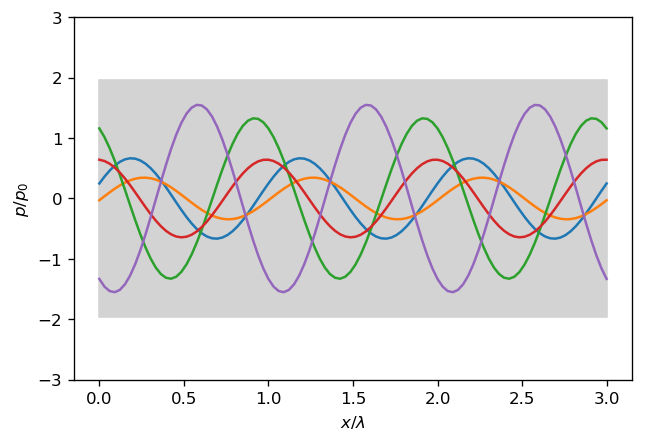}\\
(b)
\par\end{center}%
\end{minipage}\caption{Left: samples from usual GP prior with squared exponential kernel.
Right: samples from physics-informed GP prior with 1D Helmholtz kernel
\eqref{eq:1D}. 95\% confidence region for prior (gray). \label{fig:Kernels-1-1}}

\end{figure}

Before we use details of measurements we assume a \emph{prior} distribution
that allows any meaningful fields $p$. The choice of a Gaussian random
field~\eqref{eq:gp} with normalized covariance function $K(0)=1$
as a prior limits the ``allowed'' fields to magnitudes of the order
$|p|\lesssim1$ and wavelength $\lambda=2\pi/k$. In a more sophisticated
analysis expectation value and scaling of the kernel function of a
GP are optimized as hyperparameters, but here we keep them fixed to
$0$ and $1$. Fig.~\ref{fig:Kernels-1-1} shows samples from this
prior that represent possible realizations of $p$ in the absence
of information from measurements. The upper plot shows samples drawn
from a GP with a squared exponential kernel $K^{\mathrm{SE}}(x,x^{\prime})=\exp(-(x-x^{\prime})^{2}k^{2}/2)$.
The latter is often used as a covariance function if no further information
on the data is available. The shaded gray region represents the confidence
interval between $-1.96$ and $+1.96$ where $95\%$ of the data with
variance of unity lie statistically. The lower plot shows samples
from a GP prior using the 1D Helmholtz kernel $K^{\mathrm{1D}}$ \eqref{eq:1D}
that, in contrast to $K^{\mathrm{SE}}$, only allows physical fields.
When we perform a measurement of an actual field, we restrict the
possibilities to an ensemble with \emph{posterior} expectation values
and covariances. The more information we gather, the better we can
narrow down the posterior distribution for a good reconstruction of
the actual field. The reconstruction is built from the ``best guess'',
being the posterior expectation value at each point in space. In machine
learning terminology the data used to fit the GP is usually called
\emph{training} data, whereas the GP is evaluated at \emph{test} points
where the goodness of the fit can be compared to validation data.
In acoustics we are usually dealing with ``small data'', i.e. the
number of measurement points $N$ is well below 1000 and every single
data point is valuable. Therefore one will usually take all $N_{\mathrm{t}}=N$
positions for training if the speed of sound is fixed, or use $N_{\mathrm{t}}=N-1$
training point and one point for validation. The latter is known as
the ``leave-one-out'' cross validation. This validation can also
be used to check if all our microphones have been positioned and calibrated
correctly. For this purpose, we can repeat the leave-one-out method
for each microphones and check if its signal matches the GP reconstruction
from the remaining $N-1$ microphones.

\section{USING MEASURED DATA FOR REGRESSION}

\subsection{Gaussian process regression\label{subsec:GPreg}}

Here we will fit (``train'') a Gaussian process to measured data
$p(\xset_{\mathrm{t}}^{k})$ according to Ref.~\cite{Rasmussen2006}
at discrete training positions $\xset_{\mathrm{t}}^{k}$ which we
collect in the \emph{design matrix} $X_{\mathrm{t}}=(\xset_{\mathrm{t}}^{1},\v x_{\mathrm{t}}^{2},\dots,\v x_{\mathrm{t}}^{N_{\mathrm{t}}})$.
As a first step the data $p$ are normalized to be of the order $1$.
This can be achieved by normalizing $p$ by its maximum absolute value
or its empirical standard deviation. The training observations $p_{\mathrm{t}}^{k}=p(\v x_{\mathrm{t}}^{k})$
are collected in the vector $\v p_{\mathrm{t}}$. When modeled as
a Gaussian process the covariance matrix of these data is given by
$\mathrm{cov}(\v y)=K(X_{\mathrm{t}},X_{\mathrm{t}})$, where the
kernel function is evaluated for each combination of training data,
i.e. $[K(X_{\mathrm{t}},X_{\mathrm{t}})]_{jk}=K(\v x_{\mathrm{t}}^{j},\v x_{\mathrm{t}}^{k})$.
Uncertainties in measurements can be taken into account by adding
a diagonal matrix $\sigma_{\mathrm{n}}^{\,2}I$, where $\sigma_{\mathrm{n}}^{\,2}$
is called the \emph{noise variance}. Thanks to the favourable properties
of the Gaussian distribution we can write posterior expectation values
$\bar{\v p}$ and covariance matrix for $p$ evaluated at a set of
points $X=(\xset^{1},\v x^{2},\dots)$ in terms of training data at
$\xset_{\mathrm{t}}$ as
\begin{align}
\bar{\v p} & =K(X,X_{\mathrm{t}})[K(X_{\mathrm{t}},X_{\mathrm{t}})+\sigma_{\mathrm{n}}^{\,2}I]^{-1}\v p_{\mathrm{t}},\label{eq:pbar}\\
\mathrm{cov}(\v p) & =K(X,X)-K(X,X_{\mathrm{t}})[K(X_{\mathrm{t}},X_{\mathrm{t}})+\sigma_{\mathrm{n}}^{\,2}I]^{-1}K(X_{\mathrm{t}},X).\label{eq:covp}
\end{align}
The expectation values $\bar{p}_{k}=\bar{p}(\v x^{k})$ provide the
best guess for the reconstruction of $p$ by the fitted GP. Values
of the posterior variance in the diagonal entries $\sigma_{kk}^{\,2}$
of $\mathrm{cov}(\v p)$ are used to estimate posterior confidence
intervals around $\bar{p}_{k}$. For $\Delta p_{k}=1.96\sigma_{kk}$
the confidence interval $\bar{p}_{k}\pm\Delta p_{k}$ should contain
$\approx95\%$ of the possible reconstructed data and thus quantifies
the uncertainty of the reconstruction.

\subsection{Reconstruction of a 1D acoustic field\label{subsec:1Dacoustic}}

\begin{figure}
\begin{minipage}[t]{0.5\columnwidth}%
\begin{center}
\includegraphics[width=1\textwidth]{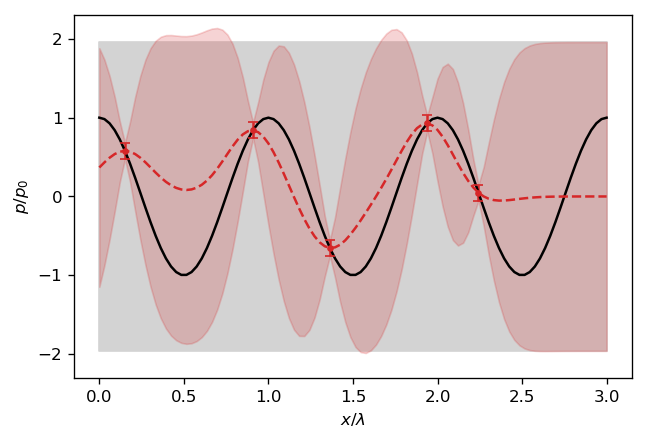}\\
(a)
\par\end{center}%
\end{minipage}%
\begin{minipage}[t]{0.5\columnwidth}%
\begin{center}
\includegraphics[width=1\textwidth]{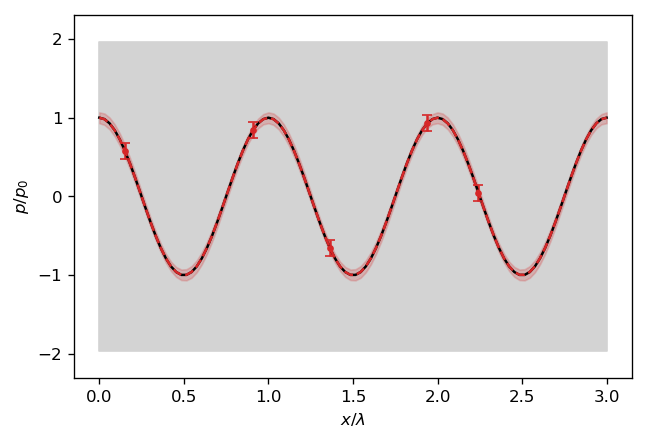}\\
(b)
\par\end{center}%
\end{minipage}\caption{1D acoustic pressure field, left: usual GP regression with squared
exponential kernel. Right: physics-informed GP regression with Helmholtz
kernel \eqref{eq:1D}. Exact data (black solid line), sampling points
(red dots with error bars) and reconstruction (red dashed line). 95\%
confidence region for prior (gray) and posterior (light red) distribution.
\label{fig:Kernels-1}\vspace{-1.5em}}
\end{figure}

As a drastic demonstration how the choice of the kernel function affects
results of a GP regression, we consider a 1D acoustic field at a single
frequency. Here physical fields are cosine functions of fixed wavelength
with scaled amplitude and shifted phase. Fig.~\ref{fig:Kernels-1}
shows results for the field sampled at 5 training points with error
bars of $\pm10\%$, i.e. a noise covariance $\sigma_{\mathrm{n}}=0.1/1.96$.
Results with a squared exponential kernel $K^{\mathrm{SE}}$ on the
left are compared to using the 1D Helmholtz kernel $K^{\mathrm{1D}}$
on the right. Relatively poor performance is realized by $K^{\mathrm{SE}}$
as it is not adapted to the physical system, while $K^{\mathrm{1D}}$
exactly fits the problem as shown in section~\ref{sec:Acoustics-with-Gaussian}.
In fact, to reproduce the 1D pressure field exactly with help of $K^{\mathrm{1D}}$,
already 2 points would suffice, so the system with 5 points is overdetermined.
If sampled data were exactly enforced without error bars this would
leads to ill conditioning and numerical instability of the matrix
inversion in Eqs.~(\ref{eq:pbar}-\ref{eq:covp}). Adding small uncertainties
that always exist in real measurements resolves this issue and is
one possible regularization method. This topic will be shortly discussed
in section~\ref{subsec:Regularization-methods}. If one looks carefully,
the posterior confidence bands match the error bars for the squared
exponential kernel, but are narrower even at the training points when
using the Helmholtz kernel. This means that part of the measurement
error in a point can be removed by taking the correlation to neighboring
measurements into account. In 2D and 3D one can expect the difference
in performance to be less extreme, since the space of solutions for
the Helmholtz equation is more diverse there. Still, the Helmholtz
kernels capture oscillating features that are absent in standard kernels
such as $K^{\mathrm{SE}}$ that don't include the system's physics.
Reconstruction of 2D and 3D fields from relatively few data will be
demonstrated in section~\ref{sec:APPLICATION-TO-ACOUSTICAL}.

\subsection{Regularization methods\label{subsec:Regularization-methods}}

Due to the possibly redundant information to fit a GP, in particular
at low frequencies, regularization is required when inverting the
covariance matrix in Eqs.~(\ref{eq:pbar}-\ref{eq:covp}). Adding
a $\sigma_{\mathrm{n}}$ term to the training data as in the 1D example
of Fig.~\ref{fig:Kernels-1} corresponds to a Tikhonov regularization,
also known as a nugget or ridge regression. In addition, or as an
alternative a Moore-Penrose pseudo-inverse representing a least-squares
solution based on a truncated singular value decomposition is suited
to restore a well-conditioned system. Especially in 2D and 3D systems
one may reduce computational load and improve the system's condition
by constructing the covariance matrix only between sensors of sufficient
distance compared to the wavelength, using the full dataset only at
the highest (Nyquist) frequency where its information is required.
Regularization should not be overused, especially in the boundary
regions of the acoustic domain, where a possibly smaller wavelength
in the adjacent structures is commonly realized. Since those structural
waves don't radiate into the acoustic domain the smoothing due to
the chosen kernel and the regularization shouldn't affect the acoustic
field very much at a certain distance.

\section{APPLICATION TO ACOUSTICAL MEASUREMENTS AND TPA\label{sec:APPLICATION-TO-ACOUSTICAL}}

\subsection{Field reconstruction and validation by nearby sensors}

\begin{figure}
\begin{minipage}[t]{0.5\columnwidth}%
\begin{center}
\includegraphics[width=0.9\textwidth]{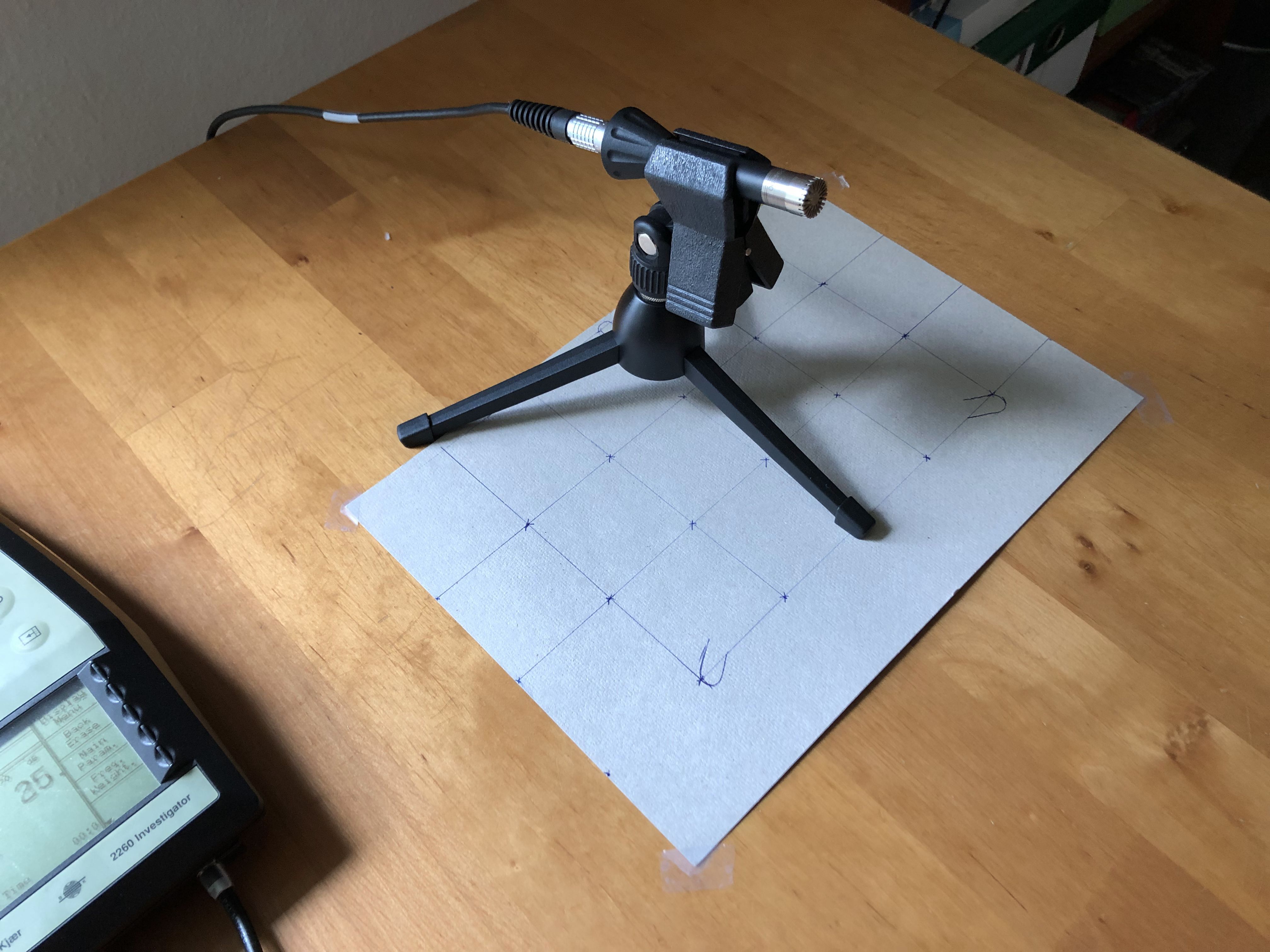}\\
(a)
\par\end{center}%
\end{minipage}\hspace*{\fill}%
\begin{minipage}[t]{0.5\columnwidth}%
\begin{center}
\includegraphics[width=1\textwidth]{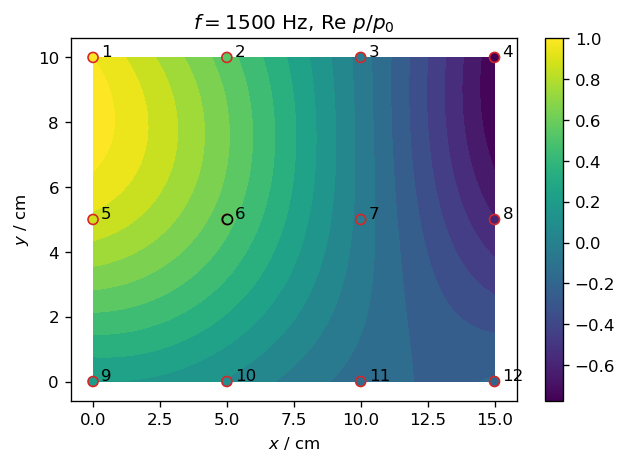}\\
(b)
\par\end{center}%
\end{minipage}

\caption{Left: Test setup of microphone moved to 12 positions in a $15\times10\,\mathrm{cm}$
rectangle on a desk. Right: Microphone positions (circles) and reconstructed
real part of $p$ at $f=1.5\,\mathrm{kHz}$. Testing point 6 marked
in black is used for validation rather than training the GP. \label{fig:Kernels-1-2-1-1-1}}

\end{figure}

Now we apply the acoustic pressure field reconstruction by GPs as
described in the previous section to a grid of microphone positions
in an actual measurement. Besides demonstrating the main features
of GP regression, this example is useful in practice to validate the
plausibility of the signal of each single transducer in a microphone
array by a GP constructed from the remaining transducers. We consider
a $3\times4$ grid of microphone positions at a spacing of $5\,\mathrm{cm}$.
Fig.~\ref{fig:Kernels-1-2-1-1-1} shows the actual setup where a
measurement microphone is moved in a plane parallel to the surface
of a wooden desk in a room without acoustic treatment. The sound field
is produced by a loudspeaker, and transfer functions $H_{pU}\propto p$
have been measured from the speaker preamp input voltage to the microphone
preamp output voltage proportional to the sound pressure $p$. GP
regression with the 3D Helmholtz kernel $K^{\mathrm{3D}}$ yields
a reconstruction of $H_{pU}(\v x)\propto p(\xset)$ at any point in
space together with uncertainty information. Here we consider only
the real part of the spectrum. Since similar results are expected
for the imaginary part, a phase-accurate reconstruction is possible
within the shown validity range. 

\begin{figure}
\begin{minipage}[t]{0.5\columnwidth}%
\begin{center}
\includegraphics[width=1\textwidth]{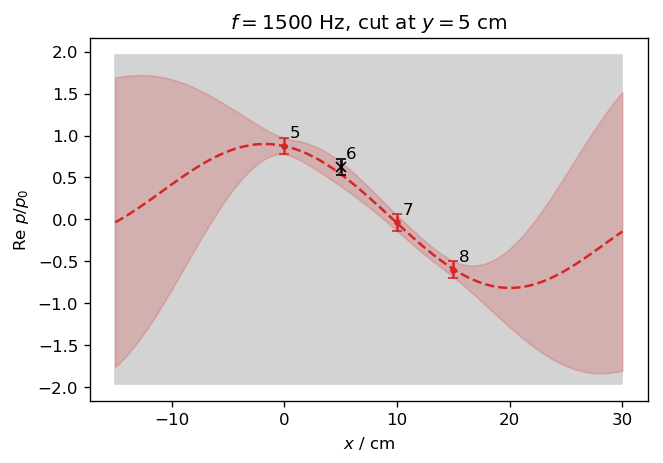}\\
(a)
\par\end{center}%
\end{minipage}\hspace*{\fill}%
\begin{minipage}[t]{0.5\columnwidth}%
\begin{center}
\includegraphics[width=1\textwidth]{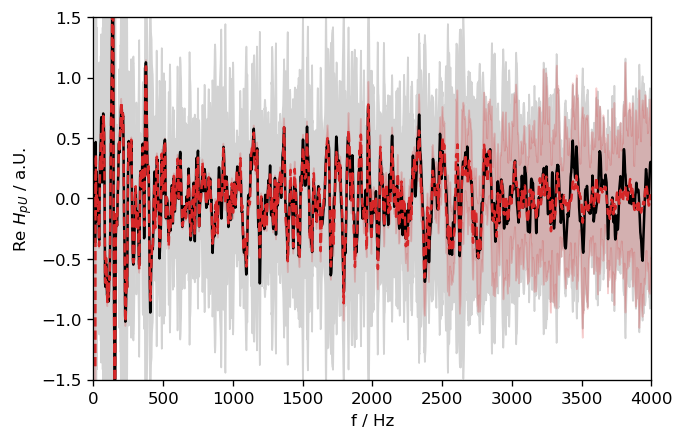}\\
(b)
\par\end{center}%
\end{minipage}

\caption{Left: Real part of $p$ field at $y=5\,\mathrm{cm}$ at $f=1.5\,\mathrm{kHz}$
tested at position 6 (black cross). Reconstruction (red dashed), 95\%
confidence bands for prior (gray) and posterior (light red). Right:
Reconstruction of the transfer function $H_{pU}$ at position 6, validation
curve as a black solid line. \label{fig:Kernels-1-2-1}\vspace{-0.5em}}
\end{figure}

Fig.~\ref{fig:Kernels-1-2-1-1-1}b shows an example of a reconstruction
of the expected value of normalized $p$ at a frequency $f=1.5\,\mathrm{kHz}$.
The according GP has been constructed from all positions except the
testing point $6$. In Fig.~\ref{fig:Kernels-1-2-1}a a cut at $y=5\,\mathrm{cm}$
is displayed, including the 95\% confidence band predicted by the
GP. The actual value of $p$ at point 6 matches the reconstruction
within the error bars. Outside the domain of measurement, the reconstruction
error grows, limiting extrapolation to a fraction of the wavelength.
This is in clear contrast to the much simpler 1D case in Fig.~\ref{fig:Kernels-1}
where a globally exact reconstruction is given from just two training
points. 

Fig.~\ref{fig:Kernels-1-2-1}b shows the reconstruction of $H_{pU}$
by the fitted GP compared to the actual signal at position 6 across
a frequency range of $10-4000\,\mathrm{Hz}$. Below $f=3\,\mathrm{kHz}$
one can observe an excellent match and nearly vanishing error bands.
Approaching the Nyquist frequency $f=3.4\,\mathrm{kHz}$ for $5\,\mathrm{cm}$
sensor distance the reconstruction becomes gradually worse and the
error bands grow to the confidence bands of the prior - the fit yields
practically no information there. For validation purposes it is important
that the sensor's signal rarely leave the confidence interval independently
from the frequency. One can take this example either as a verification
that the GP regression has worked well, or, if one already assumes
this, that the measurement at position 5 has provided physically meaningful
values in accordance with the remaining measurements.\vspace{-0.5em}

\subsection{Acoustic source strength estimation for TPA}

The presented GP regression for acoustic fields works only in source-free
domains per construction. If sources are present in the domain, we
can apply the superposition principle in the classical way to split
the Helmholtz equation into two parts:
\begin{enumerate}
\item A solution $p_{\mathrm{h}}(\v x)$ of the homogeneous equation~\eqref{eq:psf}
fulfilling the boundary conditions.
\item A particular free-field solution $p_{\mathrm{i}}(\v x)$ of the inhomogeneous
equation~\eqref{eq:p} with sources $q(\xset)$.
\end{enumerate}
For $p_{\mathrm{i}}(\xset)$ we use the free-field fundamental solution
$G(\v x,\v x^{\prime})$ for a superposition of point sources $q(\xset)=\sum_{n}q_{n}\delta(\xset-\xset_{n})$.
If the source locations are known we can estimate their respective
strength $q_{n}$ by solving the according inverse problem in the
following way.
\begin{enumerate}
\item Choose a certain combination of source strengths $\qset=(q_{n})$.
\item Subtract $p_{\mathrm{i}}(\xset^{k})=\sum_{n}q_{n}G(\v x^{k},\v x^{n})$
from the measured $p(\xset^{k})$ to obtain $p_{\mathrm{h}}(\xset^{k})=p(\v x^{k})-p_{\mathrm{i}}(\xset^{k})$.
\item Construct a GP with a Helmholtz kernel for the field $p_{\mathrm{h}}(\v x)$
by fitting training points $p_{\mathrm{h}}(\xset^{k})$.
\item Compute a loss function $\lambda$ by cross-validation of the GP,
e.g. the sum of squared distances between validation points and reconstruction
in the leave-one-out method.
\item Repeat with different $\v q$ until $\lambda$ is minimized.
\end{enumerate}
Since the GP with a Helmholtz kernel can only fit source-free fields
per design, the loss function $\lambda$ is expected to be smallest
if the data actually doesn't contain sources. This way we can find
the \textquotedbl most source-free\textquotedbl{} homogeneous solution
$p_{\mathrm{h}}$ by minimizing the loss function $\lambda(\v q)$
with respect to the source strengths $q_{k}$.

\begin{figure}
\begin{minipage}[t]{0.5\columnwidth}%
\begin{center}
\includegraphics[width=1\textwidth]{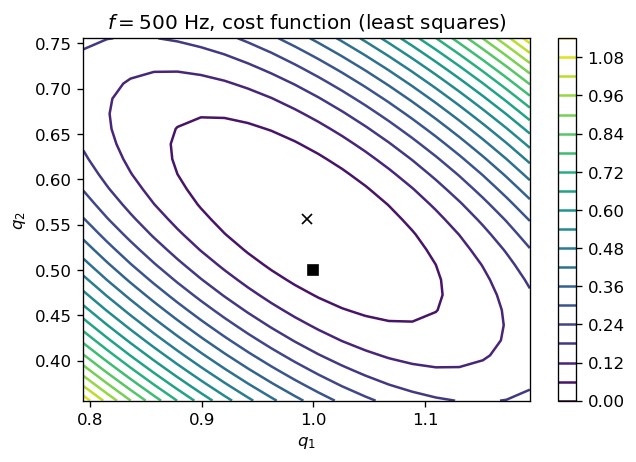}\\
(a)
\par\end{center}%
\end{minipage}\hspace*{\fill}%
\begin{minipage}[t]{0.5\columnwidth}%
\begin{center}
\includegraphics[width=1\textwidth]{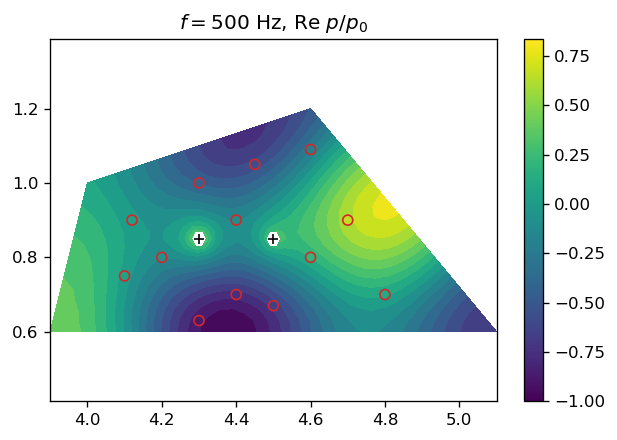}\\
(b)
\par\end{center}%
\end{minipage}

\caption{Source strength estimation via GPs for two monopoles. Left: cost function
$\lambda$ with estimated ($\times$) and actual ($\square$) solution.
Elliptical contours around a unique minimum indicate a linear problem
in $\protect\v q=(q_{1},q_{2})$. Right: microphones ($\circ$) and
sources ($+$), reconstruced $p$ field. The reconstructed field is
optically indistinguishable from the exact solution which is not plotted.
\label{fig:Kernels-1-2-1-1-2}\vspace{-1em}}
\end{figure}

Fig.~\ref{fig:Kernels-1-2-1-1-2} shows the result of this procedure
for two monopoles in a 2D cavity representing a car's engine bay.
The exact solution has been computed by an FEM simulation with acoustically
hard boundary conditions on the left and on the bottom, an impedance
condition $Z=5\rho_{0}c$ on top and a pinned $2\,\mathrm{mm}$ steel
plate on the right. Estimated strengths (normalized units) of $\v q=(0.99,0.55)$
match the exact values of $(1.0,0.5)$ well in this configuration
with 12 microphones.

It is likely that the expectation values of the estimated source strengths
from the GP approach match results from existing methods based on
fundamental solutions and optimization techniques. For acoustic TPA
such methods have the advantage that they avoid reciprocal measurement
of transfer functions to all microphone positions inside the engine
bay. One of the special features of the GP ansatz is the possibility
to propagate uncertainty information from the sensor signal up to
the final TPA results. The method is in principle suited to be used
in 3D geometry with noisy measurements, but further investigations
are required to verify this claim. The present example with just two
point-sources is not very representative for a real engine bay with
complex scattering structures. A possible next step could be the assessment
of the method's performance depending on the number of sensors and
uncertainties in their positioning within a more complex 3D geometry. 

\section*{\uppercase{Summary and Outlook}}

In this paper acoustic field reconstruction via Gaussian process regression
with kernels for physically possible fields has been introduced. The
presented examples have demonstrated potentially useful features for
application to measurement techniques within TPA. In particular, cross-validation
allows to identify inconsistencies in single sensor signals and to
estimate acoustical source strengths without prior knowledge of transfer
functions. For practical applications a number of open questions remains,
especially with respect to robustness under sensor positioning errors
and noise. In particular for source reconstruction within acoustic
TPA, non-ideal source distributions and the presence and shape of
scattering bodies have to be investigated before proceeding to practical
applications in engineering. In the future the approach could be extended
to more complex vibro-acoustic systems if elastodynamic equations
and appropriate coupling conditions were introduced. This will most
likely turn out to be a much more difficult task than modeling an
acoustic domain with known wavenumber. A representation of complex
structures by effective plate or beam models valid in a certain frequency
range could be used to yield insight into the system's overall physical
behavior. 

\section*{\vspace{-0.8em}\uppercase{Acknowledgements}}

The author would like to thank Eugène Nijman and Udo von Toussaint
for insightful discussions. This study is a contribution to the \emph{Reduced
Complexity Models} grant number ZT-I-0010 funded by the Helmholtz
Association of German Research Centers.

%\hl{AO: I have not updated the reference style yet...}

\global\long\def\refname{\normalfont\selectfont\normalsize}%

\section*{\vspace{-0.8em}\uppercase{References}}

\vspace{-18pt}

\bibliographystyle{IEEEtran}
\bibliography{Albert2019_ICA_TPA}

\end{document}